\documentclass[twocolumn,amsmath,amssymb,aps,prd,10pt,superscriptaddress,nofootinbib,noeprint,preprintnumbers,floatfix]{revtex4-1}
\pdfoutput=1

\usepackage{graphicx, color}

\usepackage[letterspace=-10]{microtype} 

\usepackage{bm, amsmath, amsfonts}
\usepackage{multirow, tabularx, dcolumn}
\usepackage{mathtools} 
\usepackage{booktabs}
\usepackage{placeins}
\usepackage{soul} 
\usepackage{subdepth} 

\usepackage[utf8]{inputenc} 
\usepackage{hyperref}

\usepackage{xcolor}
\definecolor{jlab_red}{RGB}{192,39,45}
\definecolor{jlab_orange}{RGB}{249,102,0}
\definecolor{jlab_blue}{RGB}{47,122,121}
\definecolor{jlab_green}{RGB}{65,125,10}
\definecolor{jlab_grey}{RGB}{125,125,125}

\definecolor{dstarpi_s}{RGB}{191,39,45}
\definecolor{dstarpi_d}{RGB}{65,125,10}
\definecolor{dstarpi_mix}{RGB}{248,102,0}
\definecolor{dpi_d}{RGB}{51,92,129}

\renewcommand{\star}{\ensuremath{\ast}}

\newcommand{\ccz}{\ensuremath{\chi_{c0}}}

\newcommand{\cct}{\ensuremath{\chi_{c2}}}

\newcommand{\DD}{\ensuremath{D\bar{D}}}
\newcommand{\DsDs}{\ensuremath{D_s\bar{D}_s}}
\newcommand{\DDst}{\ensuremath{D\bar{D}^\star}}
\newcommand{\DstDst}{\ensuremath{D^\star\bar{D}^\star}}
\newcommand{\DsDsst}{\ensuremath{D_s\bar{D}_s^\star}}
\newcommand{\DsstDsst}{\ensuremath{D^\star_s\bar{D}_s^\star}}

\newcommand{\psiom}{\ensuremath{\psi\omega}}

\newcommand{\etce}{\ensuremath{\eta_c\eta}}
\newcommand{\etcep}{\ensuremath{\eta_c\eta^\prime}}

\newcommand{\SLJ}[3]{\ensuremath{{\:\!}^{{#1}\!}{#2}_{#3}}}
\newcommand{\SLJc}[3]{\ensuremath{  \{  {\:\!}^{{#1}\!}{#2}_{#3}  \}  }}

\hypersetup{%
pdftitle = {Scalar and tensor charmonium resonances in coupled-channel scattering from QCD},
pdfsubject = {QCD},
pdfkeywords = {QCD, Hadron, Charm, Physics, Lattice, Meson, Scattering},
pdfauthor = {Hadron Spectrum Collaboration},
colorlinks = {true},
filecolor = {black},
linkcolor = {jlab_blue},
menucolor = {black},
citecolor = {jlab_green},
urlcolor = {jlab_green},
}{}

\begin{document}

\title{Scalar and tensor charmonium resonances in coupled-channel scattering from QCD}

\author{David~J.~Wilson}\email{d.j.wilson@damtp.cam.ac.uk}
\affiliation{DAMTP, University of Cambridge, Centre for Mathematical Sciences, Wilberforce Road, Cambridge CB3 0WA, UK}
\author{Christopher~E.~Thomas}
\email{c.e.thomas@damtp.cam.ac.uk}
\affiliation{DAMTP, University of Cambridge, Centre for Mathematical Sciences, Wilberforce Road, Cambridge CB3 0WA, UK}
\author{Jozef~J.~Dudek}
\email{dudek@jlab.org}
\affiliation{\lsstyle Thomas Jefferson National Accelerator Facility, 12000 Jefferson Avenue, Newport News, VA 23606, USA}
\affiliation{Department of Physics, College of William and Mary, Williamsburg, VA 23187, USA}
\author{Robert~G.~Edwards}
\email{edwards@jlab.org}
\affiliation{\lsstyle Thomas Jefferson National Accelerator Facility, 12000 Jefferson Avenue, Newport News, VA 23606, USA}

\collaboration{for the Hadron Spectrum Collaboration}
\noaffiliation

\date{\today}

\begin{abstract}
\noindent
We determine $J^{PC}=0^{++}$ and $2^{++}$ hadron-hadron scattering amplitudes in the charmonium energy region up to 4100 MeV using lattice QCD, a first-principles approach to QCD. Working at $m_\pi\approx 391$~MeV, more than 200 finite-volume energy levels are computed and these are used in extensions of the L\"uscher formalism to determine infinite-volume coupled-channel scattering amplitudes. We find that this energy region contains a single $\chi_{c0}$ and a single $\cct$ resonance. Both are found as pole singularities on the closest unphysical Riemann sheet, just below $4000$ MeV with widths around 70 MeV. The largest couplings are to kinematically-closed $\DstDst$ channels in $S$-wave, and couplings to several decay channels consisting of pairs of open-charm mesons are found to be large and significant in both cases. Above the ground state $\chi_{c0}$, no other scalar bound-states or near-$\DD$ threshold resonances are found, in contrast to several theoretical and experimental studies.
\end{abstract}

\maketitle

\noindent
\emph{Introduction} --- 
The experimental mapping of the spectrum of excited hadrons containing a charm-anticharm pair has seen rapid progress in recent years. Driven initially by the discovery of the $X(3872)$~\cite{Belle:2003nnu}, more novel observations quickly followed, including states with an apparent four-quark nature, such as the $Z_c(3900)$~\cite{BESIII:2013ris,Belle:2013yex}. Work to decipher this new hadron spectroscopy, going beyond the simple $c\bar{c}$ quark model, is underway~\cite{Shepherd:2016dni,Esposito:2016noz,Guo:2017jvc,Brambilla:2019esw,JPAC:2021rxu}. Within the standard model of particle physics lies Quantum Chromodynamics (QCD), the theory of interacting quarks and gluons, which describes hadrons and their interactions. While the theory is well-defined, it remains challenging to perform calculations of its spectrum owing to its strongly-coupled nature.

Charm quarks are heavy enough that relativistic effects are typically sub-leading, and models built using potentials have proven successful in describing the low-lying spectrum~\cite{Eichten:1974af,Eichten:1978tg,Richardson:1978bt,Godfrey:1985xj,Bali:2000gf,Brambilla:1999xf}. These approaches work well for states below $D\bar{D}$ threshold whose lifetimes are relatively long, with charm-anticharm annihilation and radiative transitions being the dominant modes of decay. However, above this point states can decay more rapidly to systems of open-charm mesons, and the physics of coupling to decay modes, where we treat excited states as \emph{resonances}, becomes important.

This article aims to address a key weakness in our present understanding: knowledge of resonance decays from first-principles in QCD. We compute using \emph{lattice QCD} to determine resonance masses, widths, and decay modes. We begin by considering what might na\"ively expected to be relatively simple systems: isoscalar scalar and tensor resonances in the approximation where charm-anticharm annihilation is forbidden.\footnote{This is well-defined theoretically, and well-justified empirically given the modest hadronic widths observed for states below $\DD$ threshold.} A summary of the general approach, which takes advantage of the finite spatial volume of the lattice to determine scattering amplitudes in which resonances appear, is given in a recent review~\cite{Briceno:2017max}. 
The discrete spectra extracted from corelation functions computed using lattice QCD can be translated into infinite-volume coupled-channel scattering amplitudes using the L\"uscher formalism~\cite{Luscher:1990ux} and extensions. The scattering amplitudes so obtained contain resonances as pole singularities in much the same way as experimental analyses. The pole positions yield the masses and widths, and the pole residues factorize into the channel couplings, enabling partial widths to be estimated.

In this short report, we present results for resonances found in $J^{PC}=0^{++}$ and $2^{++}$. In an accompanying longer article~\cite{Wilson:2023_long}, we give more details of our approach, and provide other amplitudes extracted in this work including $J^{PC}=3^{++}$, which is found to contain a $\chi_{c3}$ resonance, and negative parity $J^{PC}=\{1,2,3\}^{-+}$ waves which lack strong scattering, although $2^{-+}$ contains a near-threshold bound-state.\footnote{The closest previous work considering some of these channels in Lattice QCD is Ref.~\cite{Prelovsek:2020eiw}, and a comparison with the calculation reported on here can be found in the longer article, Ref.~\cite{Wilson:2023_long}.}

\emph{Computing finite-volume spectra} ---
We perform calculations using lattices with two degenerate dynamical light quark flavors and a heavier dynamical strange quark~\cite{Edwards:2008ja,Lin:2008pr}, with a light quark mass value such that $m_\pi\approx 391$~MeV.
The valence charm quarks have the same action as the light and strange quarks and are tuned to approximately reproduce the physical $\eta_c$ mass~\cite{Liu:2012ze}. 
Three volumes are employed corresponding to $L/a_s = \{16,20,24\}$, where $L$ is the spatial extent and $a_s$ is the spatial lattice spacing. Anisotropic lattices with anisotropy $\xi=a_s/a_t\approx 3.5$ are used to obtain a finer energy resolution, where $a_t$ is the temporal lattice spacing. 
In the computation of two-point correlation functions, all relevant Wick contractions, including those featuring light or strange quark annihilation, are performed efficiently using distillation~\cite{Peardon:2009gh}.

No lattice QCD study to-date has considered \emph{all} hadron-hadron channels present in this energy region, even in the simplifying limit where charm-anticharm annihilation is forbidden. In this work, we compute the complete discrete energy spectrum up to around the $\psi\phi$ threshold by using a large number of interpolating operators with fermion-bilinear ($c\bar{c}$-like) and meson-meson-like structures~\cite{Dudek:2012gj}. In particular, we construct operators resembling every relevant hadron-hadron pair with the correct quantum numbers.

While our aim is to determine the $J^{PC}=\left\{0,2\right\}^{++}$ amplitudes, the reduced symmetry of the finite cubic lattice volume means that scattering in multiple $J^P$ partial-waves contributes to the same finite-volume spectra, obtained in the irreducible representations (irreps) of the cubic group~\cite{Johnson:1982yq,Moore:2005dw}. Parity is a good quantum number for systems overall at rest, but it is not when the system has net momentum. The finite-volume of the lattice imposes quantization of momentum, $\vec{p} = \frac{2\pi}{L}(i,j,k)=[ijk]$ where $i,j,k$ are integers, and we will compute spectra for several values of total scattering system momentum.
Relevant hadron-hadron scattering combinations with $J^{PC}=\left\{0,2,3\right\}^{++}$ are shown in Table~\ref{tab:pwa:rest}, with partial-waves labelled by spectroscopic notation.\footnote{We do not aim to determine any three-hadron amplitudes, but when computing the finite-volume spectra we include operators with $\eta_c \sigma$-like and $\ccz \sigma$-like structures. The corresponding energy levels are found to be decoupled from the other levels. A complete description is given in the accompanying longer article~\cite{Wilson:2023_long}.}

\begin{table}
\def\arraystretch{1.3}
\begin{tabular}{r|clccccc}
$0^{++}$ &  & $\etce,\DD,\etcep,\DsDs,\psiom,\DstDst,\psi\phi\: \SLJc{1}{S}{0}$\\[0.5ex]
\hline
\multirow{2}{*}{$2^{++}$} &  & $\etce,\DD,{\color{jlab_grey}{\etcep,}}\DsDs\: \SLJc{1}{D}{2}; \quad \DDst, {\color{jlab_grey}{\DsDsst}}\SLJc{3}{D}{2}$\\[0.5ex]
         &  & $ \psiom,\DstDst,\psi\phi \SLJc{5}{S}{2}$\\[0.5ex]
\hline
\multirow{2}{*}{$3^{++}$} &  & $ \DDst,\psiom,\DsDsst,\psi\phi\SLJc{3}{D}{3}; \quad{\color{jlab_grey}{\eta_c \sigma \SLJc{1}{F}{3}}}$\\
         &  & $ \psiom,\DstDst, {\color{jlab_grey}{\psi\phi,\DsstDsst}}\SLJc{5}{D}{3}$\\
\end{tabular}
\caption{Hadron-hadron \SLJ{2S+1}{\ell}{J} total spin ($S$), orbital angular momentum ($\ell$) and total angular momentum combinations ($J$) present for $J^{PC}=\{0,2,3\}^{++}$ scattering. Those given in grey indicate that the corresponding operator constructions were included, but that the scattering channel was found to be decoupled or otherwise not relevant at these energies.}
\label{tab:pwa:rest}
\vspace{-5mm}
\end{table}

\begin{figure}[thb]
\includegraphics[width=0.99\columnwidth]{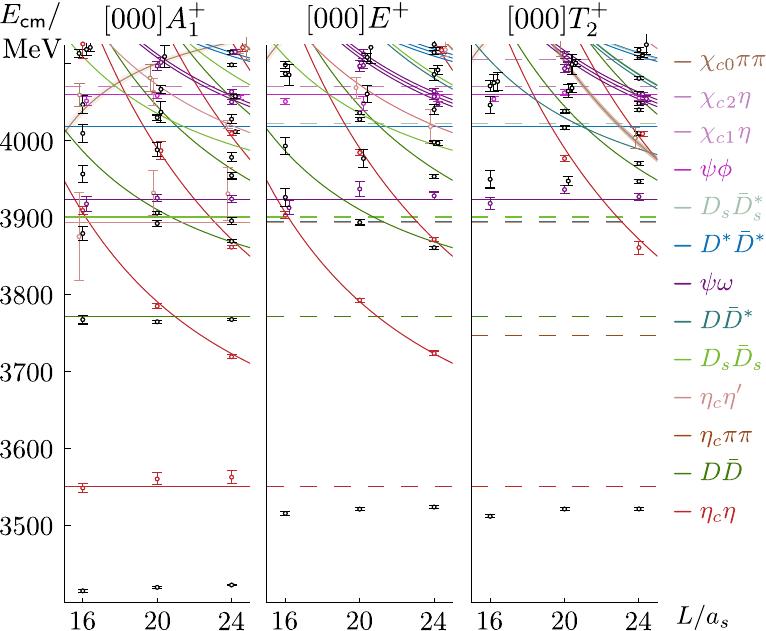}
\caption{Spectra in irreps $\Lambda^{P} = A_1^{+}$, $E^{+}$ and $T_2^{+}$ with zero overall momentum, having leading $J^{PC}=0^{++}$, $2^{++}$ and $2^{++}$ partial waves respectively.
Points are the computed finite-volume energies colored according to their dominant operator-overlap, with colors given in the key on the right. Black points have large overlap with both $c\bar{c}$-like and $D\bar{D}$-like operators.
Solid curves indicate non-interacting meson-meson energies and dashed lines indicate kinematic thresholds. Degenerate non-interacting levels are indicated by multiple parallel curves, slightly displaced in energy for visual clarity.
}
\label{fig_spec}
\vspace{-3mm}
\end{figure}


In Fig.~\ref{fig_spec} we present a selection of computed finite-volume spectra for zero overall momentum for irreps having $J^{PC}=0^{++}$ and $2^{++}$ as lowest partial-waves. Additional spectra with overall non-zero momentum and with leading $J^{PC}$=$\{1,2,3\}^{-+}$ or $3^{++}$ at zero momentum are presented in Ref.~\cite{Wilson:2023_long}. 
In each of the three panels in the figure a level is observed below $\eta_c \eta$ threshold with very little volume dependence, and these levels correspond to the stable $\ccz(1P)$ bound state (left), and the stable $\cct(1P)$ bound state (middle, right).
From $\eta_c \eta$ threshold up to around 3900 MeV there is a one-to-one correspondence between the computed energies and the levels expected in the absence of interactions, and energy shifts from these non-interacting levels are typically small, suggesting only mild interaction strength. Higher up in energy there appear to be extra levels, and more significant departures from the non-interacting spectrum, which may be due to the presence of one or more resonances. To draw more definite conclusions we must determine infinite-volume scattering amplitudes, constrained by these spectra.


\emph{Scattering amplitudes} --- 
The coupled-channel scattering $t$-matrix is obtained from finite-volume energies using L\"uscher's finite-volume quantization condition~\cite{Luscher:1990ux}, generalized for hadron-hadron scattering for hadrons with arbitrary spin~\cite{Briceno:2014oea}, 
\begin{align}
\det\Bigl[\bm{1}+i \bm{\rho}(E)\cdot\bm{t}(E)\cdot\bigl(\bm{1}+i\bm{\mathcal{M}}(E,L)\bigr)\Bigr]=0,
\label{eq_det}
\end{align}
where $\bm{t}(E)$ is the scattering $t$-matrix, $\bm{\mathcal{M}}(E,L)$ is a matrix of known functions dependent on the volume and irrep, and $\bm\rho$ is a diagonal matrix of phase-space factors, ${\rho_i=2k_i/E}$. $E$ is the centre-of-momentum frame energy and $k_i$ the momentum of each hadron in that frame for hadron-hadron channel $i$. 

The matrices in Eq.~\ref{eq_det} are in the space of relevant hadron-hadron channels and partial waves, as shown in Table~\ref{tab:pwa:rest}. Since many channels contribute, $\bm{t}(E)$ is underconstrained at any given value of $E$, and it is necessary to \emph{parameterize} the energy dependence. We make use of amplitudes of the form,
\begin{align}
\big[t^{-1}\big]_{ij} = (2k_i)^{-\ell_i}\: \big[K^{-1}\big]_{ij}\:(2k_j)^{-\ell_j} + I_{ij}\;,
\label{eq_tmat}
\end{align}
where $K_{ij}$ are the elements of a symmetric matrix that is real for real $s=E^2$.
$S$-matrix ($s$-channel) unitarity mandates that $\mathrm{Im}\, I_{ij} = -\rho_i$,
while a real part can optionally be generated through a dispersion relation as described in App. B of Ref.~\cite{Wilson:2014cna}.

\begin{figure*}
\includegraphics[width=0.99\textwidth]{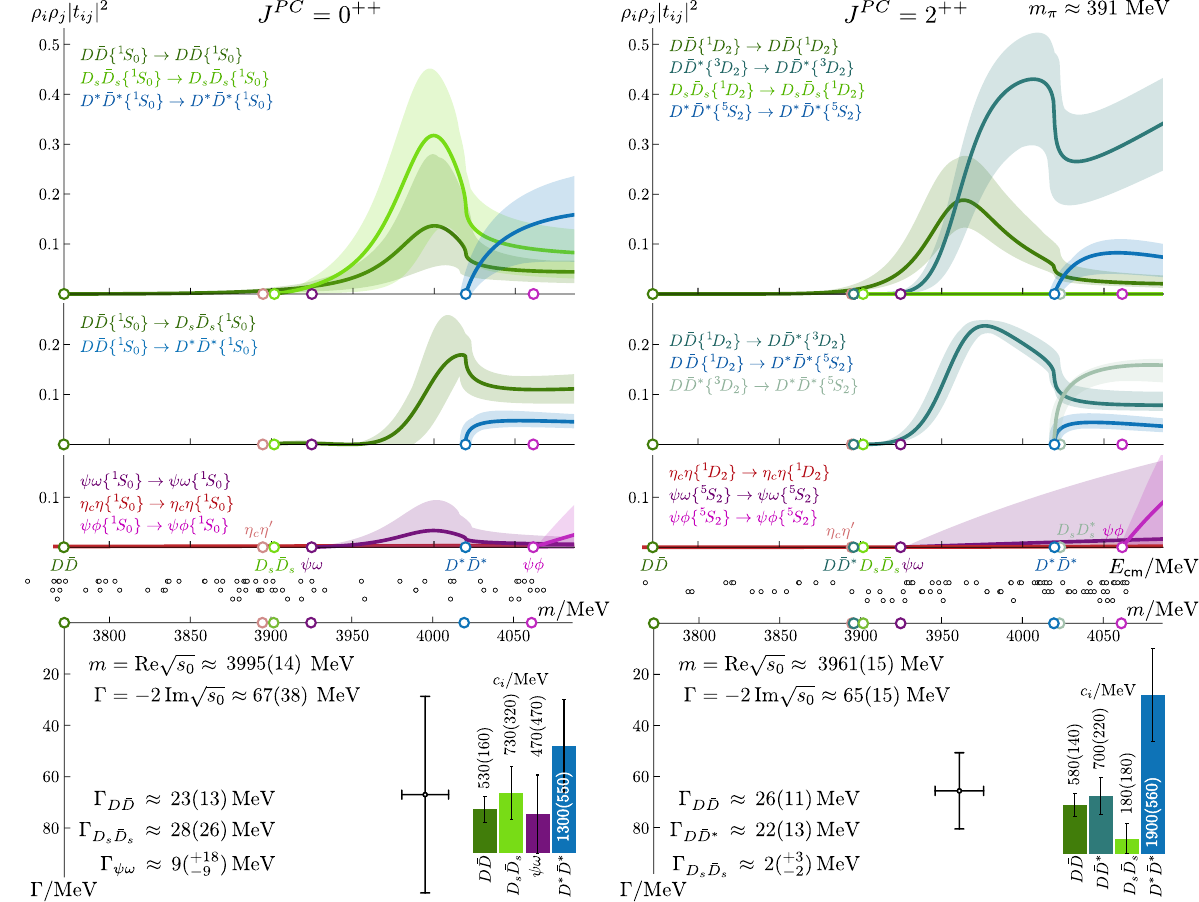}
\caption{
Scattering amplitudes (top) for $J^{PC}=0^{++}$ (left) and $2^{++}$ (right). A single representative amplitude is plotted for each $J^{PC}$ as $\rho_i\rho_j|t_{ij}|^2$, which is similar to the scattering cross section. Errorbands are determined by sampling the parameter uncertainties determined from the $\chi^2$ minimum. Small circles on the horizontal axes mark the locations of key hadron-hadron thresholds. Energies used to constrain the amplitudes (middle, open circles) and resonance poles (bottom) are also shown, with the pole parameters reflecting the full uncertainty over parameterization variation, as presented in Ref.~\cite{Wilson:2023_long}. 
}
\label{fig:amps}
\vspace{-3mm}
\end{figure*}

The $0^{++}$ and $2^{++}$ amplitudes are determined using the constraint provided by 90 and 86 energy levels respectively, taken from spectra both at-rest and for nonzero total momentum. Additional levels from other irreps are used to fix the $3^{++}$ and negative parity waves which also contribute to Eq.~\ref{eq_det}, leading to constraint from more than 200 levels in total. Finite-volume energy levels corresponding to the low-lying $\ccz(1P)$ and $\cct(1P)$ bound states are observed below $\eta_c\eta$ threshold in Fig.~\ref{fig_spec}, but because they do not constrain the amplitudes in the physical scattering region we choose not to include them when determining the scattering amplitudes.

In $J^{PC}=\{0,2,3\}^{++}$, amplitudes that prove to be capable of describing the finite-volume spectra are found to house resonance poles coupled to channels consisting of pairs of open-charm mesons. Such poles can be efficiently parameterized by including terms of form $K_{ij}= g_i g_j/(m^2-s)$, with parameters $m$ and $\{g_i\}$, and increased flexibility in the amplitude comes from adding a low-order polynomial in $s$ to this pole term.
The free parameters in the amplitudes are determined by comparing the spectrum predicted by Eq.~\ref{eq_det} for a given parameterization to the lattice QCD spectra, via a $\chi^2$ minimization~\cite{Guo:2012hv,Wilson:2014cna,Woss:2020cmp}.

To reduce bias from selection of a specific choice of form for $K_{ij}$, we consider a range of parameterizations, and when quoting properties of the scattering amplitudes such as pole positions and couplings, we take an envelope over the range of values coming from all parameterization choices that describe the spectra with reasonable $\chi^2/N_\mathrm{dof}$. Representative examples resulting from this procedure are shown in Fig.~\ref{fig:amps}. 

In both scalar and tensor cases, clear narrow peaks are visible near 4000 MeV, likely indicating resonant behavior. In the scalar case, the peaks in elastic $\DD$ and $\DsDs$ appear at the same location, and both are distorted in their high-energy tail by the opening of the $\DstDst$ channel. In the tensor case, the elastic $\DDst$ energy-dependence is sculpted by the $D$-wave threshold opening only slightly below the resonance leading to a peaking behavior at a slightly larger energy than the peak in \DD. No peak is seen in tensor \DsDs.

\emph{Poles \& Interpretation} --- 
The partial-wave $t$-matrices we use are analytic functions of $s=E^2$ apart from branch cuts opening at thresholds, and poles corresponding to bound-states and resonances.\footnote{Neither the $t$-matrices we utilize, nor the finite-volume quantization condition~\cite{Luscher:1990ux,Briceno:2014oea}, explicitly include singularities due to hadron exchange processes in the $t$ and $u$-channels, such as pion exchange in $\DDst$ and $\DstDst$, which has been highlighted recently for the related $T_{cc}(3875)^+$~\cite{Padmanath:2022cvl,Du:2023hlu}, and $NN$~\cite{Green:2021qol} scattering. Work is underway to extend the finite-volume formalism~\cite{Raposo:2023nex} to explicitly account for such physics.} 
Passing through the cuts from the real energy axis where scattering occurs, we enter ``unphysical'' Riemann sheets on which the resonance poles live. Close to a pole, \mbox{$t_{ij} \sim c_i c_j/(s_{\textrm{pole}}-s)$}, where ${s_{\textrm{pole}} = \big(m - \tfrac{i}{2} \Gamma\big)^2}$ is the location of the pole, and $c_i$ is the coupling of the pole to channel $i$, which can be related to the partial width $\Gamma_i$ for a kinematically-open channel.
For the amplitudes in the current study which describe the computed finite-volume spectra, we find resonance poles on the  ``proximal sheet'' which has $\mathrm{Im} \, k_i < 0$ for kinematically open channels and $\mathrm{Im} \, k_i>0$ for closed channels, and which is closest to physical scattering.~\footnote{The amplitudes considered also feature poles on other, more distant, sheets that are not as relevant for scattering at real energies.}

Investigating $0^{++}$, a single resonance pole with large couplings to $\DD$, $\DsDs$ and $\DstDst$ is found on the proximal sheet in every amplitude. Its location is $m \approx 3995(14)$ MeV, $\Gamma \approx 67(38)$ MeV,  and at this energy, $\DstDst$ is a closed channel, but decays are possible to $\DD$ and $\DsDs$ with branching fractions of approximately 40\% and 60\% respectively. In most amplitudes which describe the finite-volume spectra, very small couplings are found to $\psi\omega$, although in a few cases a larger value is not ruled out, with a branching fraction no larger than 40\%.

Similarly in $J^{PC}=2^{++}$, only a single resonance pole appears on the proximal sheet, with large couplings to open $\DD$, $\DDst$ (both in $D$-wave) and closed $\DstDst$ (in $S$-wave). The pole has only very small couplings to $\DsDs$, $\etce$, $\psiom$ and $\psi\phi$, and is located at $m \approx 3961(15)$ MeV, $\Gamma \approx 65(15)$ MeV. Poles and couplings are shown in Fig.~\ref{fig:amps}. 

The use of a light quark mass heavier than the physical value, and expectations of discretization effects, precludes direct comparison of our results with experiment. Nevertheless, we expect resonance properties in the current system to have even milder dependence on the light quark mass than for lighter hadrons~\cite{Nebreda:2010wv,Wilson:2015dqa,Bolton:2015psa,Wilson:2019wfr,Molina:2020qpw,Cheung:2020mql,Gayer:2021xzv,Rodas:2023twk}, so we can view previous experimental results and theoretical predictions in the context of our results.

In the energy region below about 4000~MeV, our calculation results in a state-counting consistent with $c\bar{c}$ quark-models~\cite{Eichten:1978tg,Godfrey:1985xj}, in which the lightest scalar and tensor states are $1P$ configurations, and the excited states we have observed would correspond to the $2P$ radial excitations. The resonances found in this study favor decays to open-charm $D$-meson pairs over closed-charm final states, supporting the long-standing OZI phenomenology.

The experimental $X(3872)$ observed close to $\DDst$ threshold has motivated models with attraction between the open-charm mesons mediated by pion exchange, with enough strength to provide binding. Heavy-quark spin symmetry then suggests similar effects may occur in $\DstDst$ in $S$-wave~\cite{Albaladejo:2015dsa,Baru:2016iwj}. The scalar and tensor resonance poles found in the current calculation do have large couplings to the kinematically-closed $S$-wave $\DstDst$ channel, but in both cases the attraction is apparently not large enough to produce an additional state beyond the expectations of $c\bar{c}$ excitations.

Our results suggest a single $0^{++}$ resonance that might explain both the $\ccz(3930)$~\cite{LHCb:2020pxc} and $\ccz(3960)$~\cite{LHCb:2022aki} peak structures seen in $\DD$ and $\DsDs$ final states respectively.
Claims for an additional $\ccz$ state between 3700 and 3860 MeV appear in experiment~\cite{Belle:2017egg}, lattice~\cite{Prelovsek:2020eiw},  bound hadron-molecule models~\cite{Dong:2021bvy}, $c\bar{c}+D\bar{D}$ hadron-loop dressing models~\cite{Tornqvist:1993ng,Barnes:2005pb,Pennington:2007xr,Danilkin:2010cc,Ortega:2017qmg}, and reanalyses~\cite{Guo:2012tv,Wang:2019evy,Wang:2020elp,Deineka:2021aeu} of the experimental data, although no such state is reported in recent LHCb data~\cite{LHCb:2019lnr,LHCb:2020pxc}. Our calculation shows no indication of any such additional state.

The single $2^{++}$ resonance found in this calculation decays to $\DD$ and $\DDst$, but has at most weak coupling to $\DsDs$ and closed-charm final states. This result is not in tension with the current experimental situation, where a $\cct(3930)$ has been identified in $\DD$~\cite{Belle:2005rte,BaBar:2010jfn,LHCb:2020pxc}. 
The $X(3915)$ seen in vector-vector $\psi\omega$ scattering~\cite{Belle:2009and} could be attributed to either $0^{++}$ or $2^{++}$, but our findings indicate that interactions in this channel are rather weak.

\emph{Outlook} ---
These results at $m_\pi\approx 391$ MeV suggest a state-counting in $0^{++}$ and $2^{++}$ that is not obviously different from expectations in $c\bar{c}$ pictures. To reconcile these findings with works that find additional states at physical pion masses, distant pole singularities that do not impact the current analysis would be required to move rapidly through the complex energy plane as the light quark mass is reduced. Eliminating this possibility motivates further calculations at lighter quark masses using the current techniques.

Unifying enhancements observed in different final states by identifying pole singularities in unitarity-respecting scattering amplitudes has proven essential, as clearly observed in Fig.~\ref{fig:amps} in the $\DD$ and $\DDst$ tensor amplitudes that have different peak locations and amplitude shapes but arise due to a common state. 
Experimental candidate states appear in \emph{production} processes rather than scattering, but such processes can also be described in terms of the coupled-channel scattering $t$-matrix, and are constrained by unitarity. Future lattice calculations of electroweak production processes appear to be feasible~\cite{Hansen:2012tf,Briceno:2021xlc,Radhakrishnan:2022ubg}.

Further applications of the lattice QCD approach presented in this paper will consider other near-threshold charmonia, the $X(3872)$ channel being a particularly interesting prospect.


%
\acknowledgments

We thank our colleagues within the Hadron Spectrum Collaboration (\url{www.hadspec.org}), in particular Raul Brice\~{n}o, Andrew Jackura and Arkaitz Rodas, and also acknowledge useful discussions with Igor Danilkin, Feng-Kun Guo, Christoph Hanhart, Sasa Prelovsek, Steve Sharpe and Adam Szczepaniak.
DJW acknowledges support from a Royal Society University Research Fellowship. DJW \& CET acknowledge support from the U.K. Science and Technology Facilities Council (STFC) [grant number ST/T000694/1].
JJD acknowledges support from the U.S. Department of Energy contract DE-SC0018416 at William \& Mary, and JJD \& RGE from contract DE-AC05-06OR23177, under which Jefferson Science Associates, LLC, manages and operates Jefferson Lab. 

The software codes
{\tt Chroma}~\cite{Edwards:2004sx}, {\tt QUDA}~\cite{Clark:2009wm,Babich:2010mu}, {\tt QUDA-MG}~\cite{Clark:SC2016}, {\tt QPhiX}~\cite{ISC13Phi}, {\tt MG\_PROTO}~\cite{MGProtoDownload}, {\tt QOPQDP}~\cite{Osborn:2010mb,Babich:2010qb}, and {\tt Redstar}~\cite{Chen:2023zyy} were used. 
Some software codes used in this project were developed with support from the U.S.\ Department of Energy, Office of Science, Office of Advanced Scientific Computing Research and Office of Nuclear Physics, Scientific Discovery through Advanced Computing (SciDAC) program; also acknowledged is support from the Exascale Computing Project (17-SC-20-SC), a collaborative effort of the U.S.\ Department of Energy Office of Science and the National Nuclear Security Administration.

This work used the Cambridge Service for Data Driven Discovery (CSD3), part of which is operated by the University of Cambridge Research Computing Service (www.csd3.cam.ac.uk) on behalf of the STFC DiRAC HPC Facility (www.dirac.ac.uk). The DiRAC component of CSD3 was funded by BEIS capital funding via STFC capital grants ST/P002307/1 and ST/R002452/1 and STFC operations grant ST/R00689X/1. Other components were provided by Dell EMC and Intel using Tier-2 funding from the Engineering and Physical Sciences Research Council (capital grant EP/P020259/1). This work also used the earlier DiRAC Data Analytic system at the University of Cambridge. This equipment was funded by BIS National E-infrastructure capital grant (ST/K001590/1), STFC capital grants ST/H008861/1 and ST/H00887X/1, and STFC DiRAC Operations grant ST/K00333X/1. DiRAC is part of the National E-Infrastructure.
This work also used clusters at Jefferson Laboratory under the USQCD Initiative and the LQCD ARRA project.

Propagators and gauge configurations used in this project were generated using DiRAC facilities, at Jefferson Lab, and on the Wilkes GPU cluster at the University of Cambridge High Performance Computing Service, provided by Dell Inc., NVIDIA and Mellanox, and part funded by STFC with industrial sponsorship from Rolls Royce and Mitsubishi Heavy Industries. Also used was an award of computer time provided by the U.S.\ Department of Energy INCITE program and supported in part under an ALCC award, and resources at: the Oak Ridge Leadership Computing Facility, which is a DOE Office of Science User Facility supported under Contract DE-AC05-00OR22725; the National Energy Research Scientific Computing Center (NERSC), a U.S.\ Department of Energy Office of Science User Facility located at Lawrence Berkeley National Laboratory, operated under Contract No. DE-AC02-05CH11231; the Texas Advanced Computing Center (TACC) at The University of Texas at Austin; the Extreme Science and Engineering Discovery Environment (XSEDE), which is supported by National Science Foundation Grant No. ACI-1548562; and part of the Blue Waters sustained-petascale computing project, which is supported by the National Science Foundation (awards OCI-0725070 and ACI-1238993) and the state of Illinois. Blue Waters is a joint effort of the University of Illinois at Urbana-Champaign and its National Center for Supercomputing Applications.


\bibliographystyle{apsrev4-1}
\bibliography{biblio.bib}


\end{document}